# Timing Aware Dummy Metal Fill Methodology

Luis Charre
University of Siena
Italy

Bruno Gravano
University of Siena
Italy

Rémi Pôssas
University of Liege
Belgium

Chen Zheng
Intel
USA

**Abstract**
In this paper, we analyzed parasitic coupling capacitance coming from dummy metal fill and its impact on timing. Based on the modeling, we proposed two approaches to minimize the timing impact from dummy metal fill. The first approach applies more spacing between critical nets and metal fill, while the second approach leverages the shielding effects of reference nets. Experimental results show consistent improvement compared to traditional metal fill method.

## 1. Introduction
In modern semiconductor chip manufacturing, several process steps are required to ensure yield and reliability quality. One of such step is chemical-mechanical planarization (CMP) [1]. CMP requires certain close-to-uniform metal density across the entire chip area. For example, if metal density over one chip window is too low, then the metal shapes get over polished and results in CMP dishing defects [2]. Due to CMP defects, chip reliability is affected significantly, such as electromigration [3], etc. To avoid this kind of defect, during physical design, dummy metal fill is needed to fill dummy metal shapes into empty areas to achieve close-to-uniform metal density across the chip area [4]. The metal fill procedure is purely physical mechanism, and previously designers overlook other potential impacts caused by metal fill [5]. There are many existing research works on metal fill methodology: in [6], S. Gaskill studied how to suppress noise during metal fill; in [7], V. Suresh proposed a lithography aware metal fill method; In [8], V. Shilimkar discussed modeling of metal fill parasitic capacitance. However, previous works did not take timing impact into account.

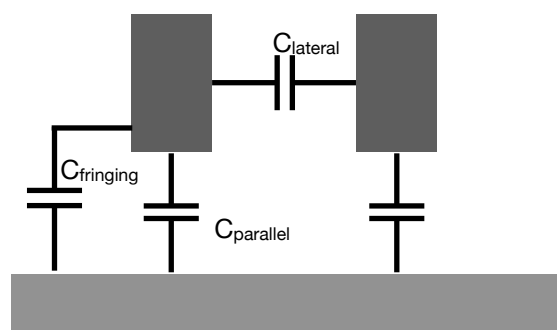

Fig 1. Parasitic coupling capacitance

## 2. Motivation
In our research, we detect that due to excessive parasitic coupling traditional blind metal fill in some cases, will significantly degrade timing on critical nets or paths. The resulting degradation can hurt chip performance and designer will be forced to reduce chip frequency to accommodate the impact. To overcome this limitation, we propose two approaches that can address this issue and achieve both yield/reliability requirement as well as maintain timing or even improved timing result.

## 3. Delay Modeling

To analyze the timing impact from metal fill, we refer to the parasitic capacitance model [9]. In Figure 1, $C_{lateral}$ is the coupling capacitance between metal wires on same layer, $C_{parallel}$ is the coupling capacitance between metal wires on adjacent layers, and $C_{fringing}$ is the coupling capacitance between side wall of a metal wire and the top/bottom surface of another metal wire on adjacent layer. Each capacitance is given by equations (1-3), the total coupling capacitance is the summation of all types of parasitic capacitances.

$$C_{lateral} = \frac{H\varepsilon_{di}}{s_{di}} \quad (1)$$

$$C_{parallel} = \frac{w\varepsilon_{di}}{t_{di}} \quad (2)$$

$$C_{fringing} = \frac{2\pi\varepsilon_{di}}{\log(t_{di}/H)} \quad (3)$$

It can be observed that the value of coupling capacitance is reversely proportional to the spacing between lateral wires and parallel wires.

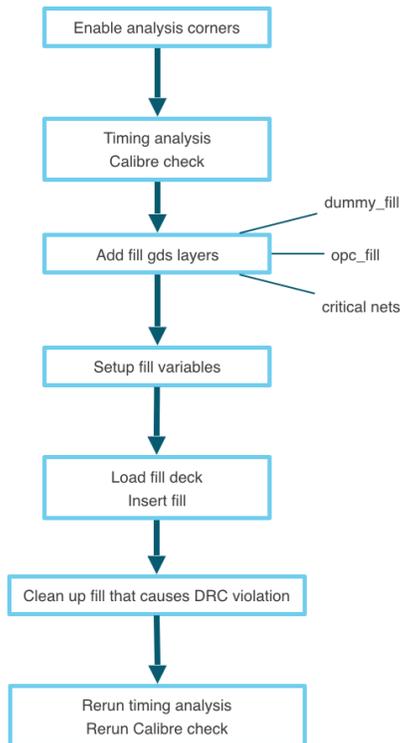

**Fig 2. Timing-driven metal fill**

Thus, keeping spacing between critical nets and dummy metal fill will reduce the coupling capacitance.

## 4. Timing-Driven metal fill

Based on the coupling capacitance model we analyzed in section 3, we propose two approaches to address the issue coming from metal fill:

*(1) increased spacing between critical nets and dummy fill;*

*(2) connect dummy fill adjacent to critical nets to reference net (e.g. ground or power net) to serve as shielding.*

Approach (1) is straightforward to understand as equations (1-3) clearly shows its dependency on spacing. It can also be easily implemented as a non-default-rule (NDR) style metal fill. However, leaving space between metal fill and critical routing nets can reduce the metal density, and potentially cause metal density violation. Shielding has been proven to be an effective approach to decrease coupling capacitance [10]. It should be noted that with recent power net methodology [11], it increases the difficulty of connecting shield to reference nets, but in our case, the metal fill is able to use whatever resources are left over, so impact will be minimized. The connection also needs to be aware of any customized routing optimizations [12]. By providing reference nets as shielding, it also helps reduce variation effects and improves reliability of given nets [13]. By improving the timing on critical nets, the timing-driven metal fill can also potentially help overall chip reliability [14]. The reduced coupling capacitance also results in less overshoot current [15][16].

It should be noted that the two approaches are mutually exclusive. However, to combine those two approaches, we developed a greedy algorithm that first globally apply approach (1) across the whole design, then design windows that violates metal density rule are identified and approach (2) is further applied. Figure 2 illustrates the flow chart of our timing-driven metal fill methodology.

## 5. Experiments

We used two blocks from the open source design or1200_fcmp and or1200_genpc [17] as our benchmark testcases. The physical synthesis is done using Synopsis ICC2 [18], and the metal fill is done through Mentor Graphics Calibre [19].

Table 1 and 2 shows the timing differences for before metal fill, after regular metal fill and after timing-driven metal fill.

**Table 1. or1200_fcmp**

| or1200_fcmp | WNS | TNS | #violation |
|---|---|---|---|
| before | -0.015 | -0.055 | 7 |
| regular | -0.016 | -0.046 | 5 |
| timing-driven | -0.024 | -0.133 | 11 |

**Table 2. or1200_genpc**

| or1200_genpc | WNS | TNS | #violation |
|---|---|---|---|
| before | -0.019 | -0.073 | 10 |
| regular | -0.019 | -0.076 | 11 |
| timing-driven | -0.031 | -0.124 | 16 |

# 6. Conclusions

In this paper, we investigated the limitations of traditional metal fill methodology. The analysis shows potential timing degradation due to impact of dummy metal fill. We proposed two approaches for reducing the coupling capacitance while maintaining the chip yield and manufacturability. Results show promising improvement from our timing-driven metal fill methodology. In the future technology nodes, the timing-driven metal fill can further take into accounts various effects, such as temperature [20] and variation [21], to provide more robust solutions.